\documentclass[10pt,sigplan,anonymous=false,balance=false]{acmart}
\renewcommand\footnotetextcopyrightpermission[1]{}
\acmConference[IWLS'22]{}{July 18-21, 2022}{Virtual Conference}

\usepackage{algorithmic}
\usepackage{tikz}
\usepackage{cleveref}
\usepackage{subcaption} 
\usepackage[shortlabels]{enumitem} 
\usetikzlibrary{calc}

\usepackage{listings}
\begin{document}

\title{An Automated Testing and Debugging Toolkit for Gate-Level Logic Synthesis Applications}

\author{Siang-Yun Lee}
\affiliation{
  \institution{LSI, EPFL}
  \country{Switzerland}
}

\author{Heinz Riener}
\affiliation{
  \institution{Cadence Design Systems}
  \country{Germany}
}

\author{Giovanni De Micheli}
\affiliation{
  \institution{LSI, EPFL}
  \country{Switzerland}
}

\begin{abstract}
Correctness and robustness are essential for logic synthesis applications, but they are often only tested with a limited set of benchmarks. Moreover, when the application fails on a large benchmark, the debugging process may be tedious and time-consuming. In some fields such as compiler construction, automatic testing and debugging tools are well-developed to support developers and provide minimal guarantees on program quality. In this paper, we adapt fuzz testing and delta debugging techniques and specialize them for gate-level netlists commonly used in logic synthesis. Our toolkit improves over similar tools specialized for the AIGER format by supporting other gate-level netlist formats and by allowing a tight integration to provide $10\times$ speed-up. 
Experimental results show that our fuzzer captures defects in mockturtle, ABC, and LSOracle with $10\times$ smaller testcases and our testcase minimizer extracts minimal failure-inducing cores using $2\times$ fewer oracle calls.
\end{abstract}
\maketitle

\section{Introduction}\label{sec:intro}
Logic synthesis is the task of turning an abstract specification into a gate-level netlist composed of logic gates while considering cost functions for area, delay, and power. Logic synthesis plays a crucial role in modern microchip design flows. Sophisticated algorithmic solutions are readily available as open-sourced libraries and tools~\cite{Soeken18,BraytonM10,NetoAT+19}. These algorithms can optimize netlists with millions of gates within only a few seconds, relying on bit-precise reasoning engines.  The inherent complexity of these engines, optimized for many corner cases, makes logic synthesis algorithms susceptible to design and implementation errors.  Moreover, algorithms are often only tested on fixed benchmark suites, such as the EPFL logic synthesis benchmarks~\cite{Amaru15benchmark}.  Due to numerous possibilities to implement the same Boolean function with different circuit structures, it is not rare that subtle faults slip through the development process and only show themselves when the algorithm is used in practice.

Motivated by the success of automated testing methods, we argue that directed testing approaches and bug-pointing tools specialized for logic synthesis applications can support the developers in detecting bugs earlier, can make implementations more robust, and ultimately lead to a reduction in the time and effort spend for debugging.  Due to the large state space and homogeneity of the commonly used netlist formats, general-purpose testing and debugging tools often are incapable of providing the necessary performance to efficiently test implementations of logic synthesis algorithms. The C++ logic network library \emph{mockturtle}~\cite{Soeken18,RienerTH+19} has deployed a framework for unit testing, continuous integration on various operating systems and compilers, and a static code analysis engine controlled by user-defined queries to aid developers.

In this paper, we present the latest additions to \emph{mockturtle} for testing and debugging gate-level netlists: (1) a fuzz tester that repeatedly generates small- and intermediate-sized netlists to hunt for bugs and (2) a testcase minimizer to isolate the failure-inducing core of potentially lengthy bug report. The two methods advance existing ones in the AIGER utilities\footnote{\url{https://github.com/arminbiere/aiger}} with the following key highlights:
\begin{enumerate}[leftmargin=*]
\item Our methods are agnostic of the network type and support different gate-level netlist formats.  Although conceptually not hard to implement, to our knowledge this is the first time that automated debugging techniques are available for logic representations such as \emph{Majority-Inverter Graphs}~(MIGs) or \emph{Xor-And-Graphs}~(XAGs). We demonstrate with a case study in \Cref{subsec:case-fuzz} that testing with more compact representations like XAGs increases the possibility of capturing rare defects.
\item Our implementations are tightly integrated into \emph{mockturtle}, which eliminates interfacing overheads and provides about $10\times$ speed-up over using external testing and debugging solutions. 
\item Our fuzz tester provides systematic approaches to test on small circuit topologies in addition to purely-random networks. Experimental results show that our topology-based fuzzer captures defects in \textit{ABC}, \textit{mockturtle} and \textit{LSOracle} using $93\%$ smaller testcases comparing to an existing AIG fuzzer \texttt{aigfuzz}$^1$. 
\item Our testcase minimizer guarantees to isolate a minimal failure-inducing core and reduces testcases more efficiently by adopting specialized structural reduction rules for gate-level netlists. Experimental results show that our minimizer isolates smaller or equal-sized cores using $50\%$ oracle calls and $50\%$ runtime comparing to an existing AIG delta debugger \texttt{aigdd}$^1$.
\end{enumerate}

\section{Background}
\subsection{Scope and Terminologies}
\emph{Logic networks} are technology-independent representations of digital circuits. They model combinational parts of gate-level circuits with directed acyclic graphs, where vertices, or \emph{nodes}, represent logic gates and edges represent interconnecting wires.
Prominent examples of logic networks include \emph{And-Inverter Graph}~(AIG)~\cite{KuehlmannPKG02}, \emph{Xor-And-Inverter Graph}~(XAG) \cite{HalecekFS17}, and \emph{Majority-Inverter Graph}~(MIG)~\cite{Amaru15}. Common terms related to structural properties of logic networks, such as \emph{primary input}~(PI), \emph{primary output}~(PO), \emph{transitive fanin cone}~(TFI), \emph{transitive fanout cone}~(TFO), and \emph{maximum fanout-free cone}~(MFFC), are defined the same as in the literature.~\cite{brayton2006scalable}

This paper focuses on testing and debugging software applications, referred to as the \emph{application under tests}~(AUTs), that take a logic network, called a \emph{testcase}, as an input. Prominent examples of such applications include implementations of logic synthesis algorithms such as rewriting~\cite{MishchenkoCB06}, resubstitution~\cite{MishchenkoBJJ11} and technology mapping~\cite{CalvinoRR0M22}. Methods to verify the correctness of the results, referred to as the \emph{verification}, are assumed to be provided. They may come from several sources:
\begin{itemize}
    \item Assertions within the program.
    \item Memory protection processes in the operating system checking for illegal memory access (typically raising segmentation faults). 
    \item \emph{Combinational equivalence checking}~(CEC)~\cite{MishchenkoCBE06} of the output network against the input testcase (for logic optimization algorithms).
    \item Additional code checking coherence of the program's internal data structures, such as checking if the network is acyclic and checking the correctness of reference counts, etc.
    \item Another algorithm of the same purpose used to provide reference solutions (for problems having an unique correct solution).
\end{itemize}
A failing verification, e.g., a non-equivalent CEC result, indicates that a \emph{defect} of the AUT is observed and the testcase used is said to be \emph{failure-inducing}. The AUT combined with its verification is referred to as an \emph{oracle}, and running the oracle with a testcase is an \emph{oracle call}.

\subsection{Fuzz Testing}
\emph{Fuzz testing}~\cite{MillerFS90} is a software testing technique heavily used to detect security-related vulnerabilities and reliability issues. It is conceptually simple, yet empirically powerful. A fuzzing algorithm involves repeatedly generating testcases and using them to test the AUT. The idea of fuzz testing first appeared in 1990, when spurious characters in the command line caused by a noisy dial-up connection to a workstation led to, surprisingly, crashes of the operating system.~\cite{MillerFS90} Nowadays, the generation of testcases in fuzz testing algorithms often involves randomness, and the testcases are supposed to be beyond the expectation of the AUT. 

Various taxonomies of fuzz testing algorithms have been developed. For example, black-box fuzzers~\cite{LeeYLSYP17} treat the AUT as a black-box oracle and only observe its input/output behavior, whereas white-box fuzzers~\cite{GodefroidLM08,CadarDE08} analyze some internal information of the AUT and generate testcases accordingly. Depending on the targeted types of AUTs, some fuzzers generate testcases based on predefined models or grammars~\cite{DeweyRH14}, whereas some other fuzzers mutate an initial \emph{seed} testcase to generate more testcases~\cite{ChaWB15}. There are often some parameters to be set for the testcase generators. A series of fuzz-tests using testcases generated with a specific parameter configuration is called a \emph{fuzz testing campaign}.~\cite{ManesHHCESW21}

\subsection{Delta Debugging and Testcase Minimization}
Given two versions of the code of a program, where the first version works but the second fails, \emph{delta debugging}~\cite{Zeller99} is a method originally proposed to extract a minimal set of changes (differences in the two versions of code) that causes the failure. The algorithm was later extended for minimizing failure-inducing testcases.~\cite{ZellerH02}

The basic idea of delta debugging is binary searching and dividing the set of components, may it be the \emph{delta} between two versions of code or the input testcase to a program, testing the program with the reduced set, keeping the subsets that preserve the failure, and increasing the granularity of division. The delta debugging algorithm (\textit{ddmin}) guarantees to find a $1$-minimal subset and requires, in the worst case, $n^2 + 3n$ oracle calls, where $n$ is the size of the given set.~\cite{ZellerH02}

Besides delta debugging being a generic method for testcase minimization, researchers have claimed that domain-specific testcase minimization techniques are more effective and efficient for some applications such as tree-structured inputs~\cite{Misherghi06}, compilers~\cite{RegehrCCEEY12} and SMT solvers~\cite{KremerNP20}.
Various open-source implementations of testcase minimization tools exist, including the general-purposed \texttt{delta}\footnote{\url{https://github.com/dsw/delta}}, \texttt{aigdd}$^1$ for the AIGER format, \texttt{ddSMT}\footnote{\url{http://fmv.jku.at/ddsmt/}} for the SMT-LIB v2 format, and the LLVM \texttt{bugpoint} tool\footnote{\url{https://llvm.org/docs/Bugpoint.html}}. Inspired by delta debugging, in this paper, we aim at providing such an effective testcase minimization tool specialized for logic networks but not limited to AIGs.

\section{Testing and Debugging Toolkit}
\subsection{Testcase Generation}\label{subsec:generation}
We develop a fuzz testing framework for testing any application that takes a logic network as input. The AUT and the verification checks are provided as a combined oracle call, thus categorizing it as a black-box fuzzer. Although in some cases of fuzzing, testing with malformed testcases is key to test the robustness of the AUT, this is not the case for our usage. In logic synthesis applications, detecting and rejecting malformed inputs, e.g. a cyclic network, are usually dealt by the parsers instead of the logic synthesis algorithms. Nevertheless, as logic synthesis applications are often only tested with some common benchmark suites, our fuzzing framework still tests them with a larger input space beyond what they are usually tested with.

To generate random testcases, we propose three parameterized methods. These methods apply to any type of network having a finite set of possible gate types. 

\smallskip

\textit{Random}: Randomly generate nodes in topological order. This method is parameterized by the starting number of PIs $n_0$, the starting number of gates $m_0$, the number of networks~$k$ of the same configuration to generate, the increment of the number of PIs $\Delta n$ and of the number of gates $\Delta m$. The generator starts from generating networks of $n = n_0$ PIs and $m = m_0$ gates and keeps a counter of how many networks have been generated. After generating $k$ networks, the values of $n$ and $m$ are increased by $\Delta n$ and $\Delta m$, respectively. Given the current values of $n$ and $m$, a network is generated by:
    \begin{enumerate}
        \item Create $n$ PIs.
        \item Randomly decide on a gate type. Assume that the type requires $\kappa$ fanins.
        \item Randomly sample $\kappa$ nodes (PIs or gates) that have been created.
        \item Randomly decide for each fanin if it is complemented.
        \item Create the gate. Repeat from step $2$ if the number of gates is smaller than $m$.
        \item Assign all nodes without fanout to be POs.
    \end{enumerate}
For network types with trivial-case simplifications (e.g., in AIGs, attempting to create an AND gate with identical fanins results in returning the fanin without creating a gate) and structural hashing enabled, the number of gates may not increase after step $4$. Thus, the loop of steps $2$ to $4$ may iterate more than $m$ times and the terminating condition is when the actual number of gates is $m$. If the parameters are set improperly, e.g., if $n=1$, this might lead to an infinite loop.

\smallskip

\textit{Topology}: Exhaustively enumerate all small-sized DAG topologies and randomly concretize them. This method is parameterized by the starting number of gates $m_0$, the lower $r_l$ and upper $r_h$ bounds on the PI-to-input ratio and the number of networks $k$ of the same configuration to generate. Upon initialization, the generator enumerates all isomorphic DAG topologies of $m = m_0$ vertices using an algorithm implemented in~\cite{HaaswijkSMM20} and randomly shuffles them. Then, it starts from generating networks of the first topology and keeps a counter of how many networks have been generated. After generating $k$ networks, the generator moves on to generating the next topology. After all topologies have been used to generate $k$ networks, the value of $m$ is incremented by $1$ and topologies of the increased size are enumerated.
Given a topology, which is specified by a DAG $G$ with hanging inputs (i.e., the topology specifies how gates are connected to each other, but not how they are connected to PIs), a random network is concretized by:
    \begin{enumerate}
        \item Let $i$ be the number of hanging inputs in $G$. Randomly decide on an integer $n$ such that $r_l \cdot i \leq n \leq r_h \cdot i$. Create $n$ PIs.
        \item For each input of $G$, randomly decide on a PI to connect to.
        \item For each vertex in $G$, randomly decide on a gate type.
        \item For each edge in $G$, randomly decide whether it is complemented.
        \item Assign the last gate to be a PO.
    \end{enumerate}
In step $1$, lower values of $n/i$ leads to higher probability that the generated network reconverges on PIs, whereas higher values of $n/i$ leads to higher probability to generate a tree-like network. The generated networks are always single output.

\smallskip

\textit{Composed}: Randomly compose a few small-sized DAG topologies to form a larger network. This method is parameterized by the lower $m_l$ and upper $m_h$ bounds of the size of DAG topologies, the starting number of components $c_0$, the starting number of PIs $n_0$, the number of networks $k$ of the same configuration to generate, the increment of the number of PIs $\Delta n$ and of the number of components $\Delta c$. Upon initialization, the generator enumerates all isomorphic DAG topologies of $m_l$ to $m_h$ vertices. Then, it starts from generating networks of $n = n_0$ PIs and composed of $c = c_0$ components and keeps a counter of how many networks have been generated. After generating $k$ networks, the values of $n$ and $c$ are increased by $\Delta n$ and $\Delta c$, respectively. Given the current values of $n$ and $c$, a network is generated by:
    \begin{enumerate}
        \item Create $n$ PIs.
        \item Randomly choose a topology $G$ from the list.
        \item For each hanging input of $G$, randomly decide on an existing node (a PI or a node in a created component) to connect to.
        \item For each vertex in $G$, randomly decide on a gate type.
        \item For each edge in $G$, randomly decide whether it is complemented.
        \item If the number of created components is smaller than $c$, repeat from step $2$.
        \item Assign all nodes without fanout to be POs.
    \end{enumerate}


\subsection{Testcase Minimization}\label{subsec:minimization}
\begin{figure}
\centering
\begin{subfigure}[t]{0.23\textwidth}
    \centering
    \includegraphics[width=\textwidth]{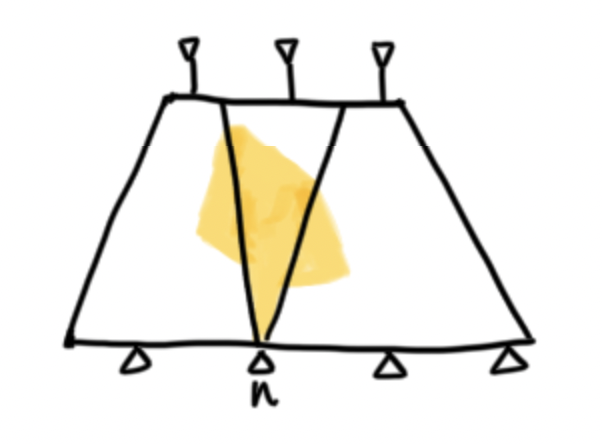}
    \caption{\textit{Remove PI}: The TFO of $n$ is simplified.}\label{subfig:remove_PI}
\end{subfigure}
\hfill
\begin{subfigure}[t]{0.23\textwidth}
    \centering
    \includegraphics[width=0.8\textwidth]{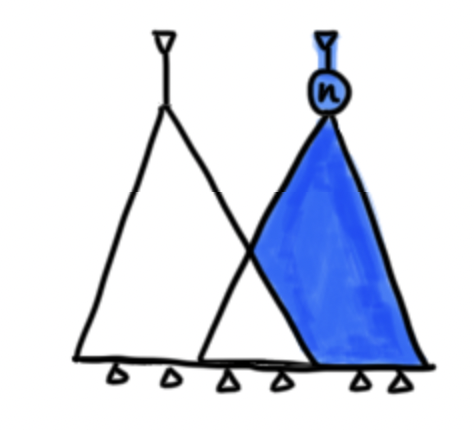}
    \caption{\textit{Remove PO}: The MFFC of $n$ is removed.}\label{subfig:remove_PO}
\end{subfigure}
\hfill
\begin{subfigure}[t]{0.23\textwidth}
    \centering
    \includegraphics[width=\textwidth]{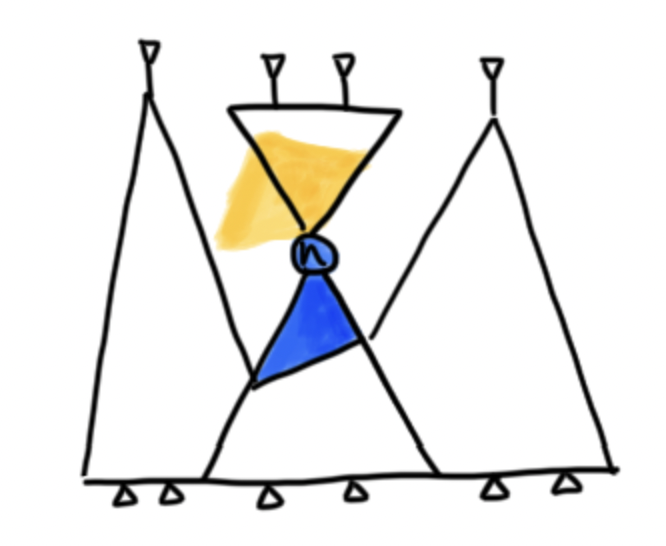}
    \caption{\textit{Substitute gate}: The MFFC of $n$ is removed and the TFO of $n$ is simplified.}\label{subfig:substitute_gate}
\end{subfigure}
\hfill
\begin{subfigure}[t]{0.23\textwidth}
    \centering
    \includegraphics[width=0.7\textwidth]{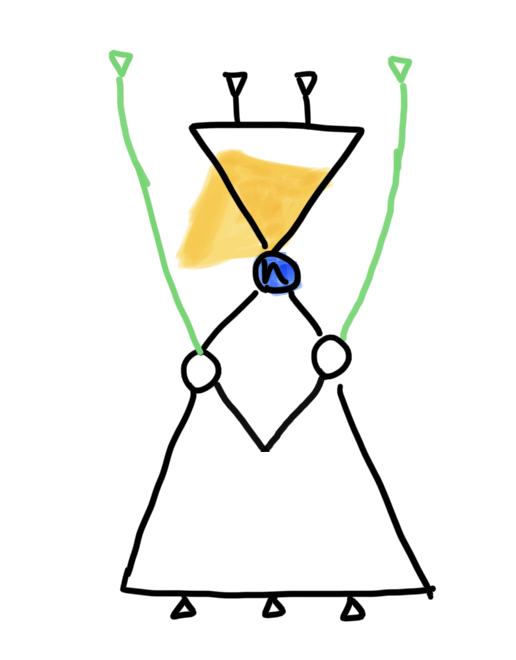}
    \caption{\textit{Simplify TFO}: The TFO of $n$ is simplified.}\label{subfig:simplify_TFO}
\end{subfigure}
\hfill
\begin{subfigure}[t]{0.23\textwidth}
    \centering
    \includegraphics[width=\textwidth]{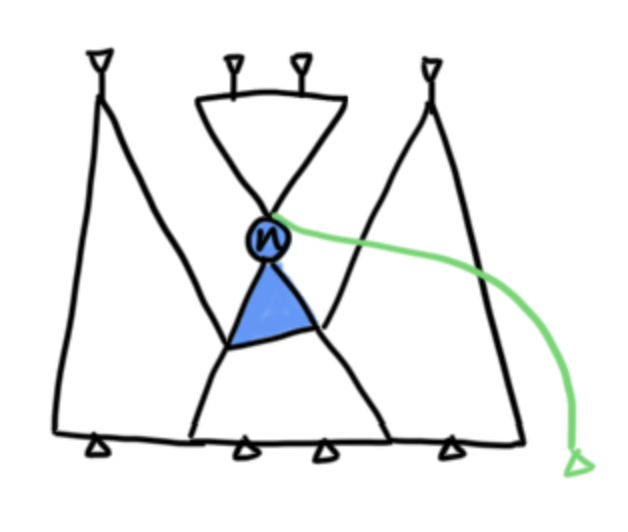}
    \caption{\textit{Remove MFFC}: The MFFC of $n$ is removed.}\label{subfig:remove_MFFC}
\end{subfigure}
\hfill
\begin{subfigure}[t]{0.23\textwidth}
    \centering
    \includegraphics[width=0.9\textwidth]{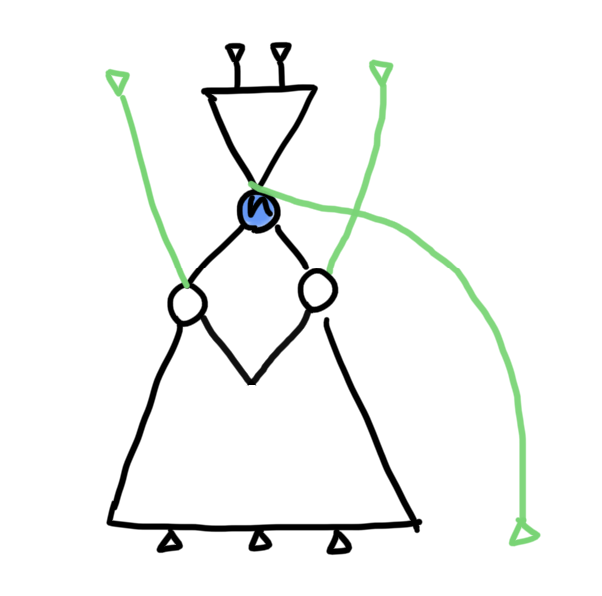}
    \caption{\textit{Remove gate}: Only $n$ is removed.}\label{subfig:remove_gate}
\end{subfigure}
\hfill
\vspace{-1em}
\caption{Illustration of the reduction stages.}
\label{fig:stages}
\end{figure}

Assuming that the concerned defect is deterministic, there is a \emph{core} in any given failure-inducing testcase, which is a subset of the testcase essential for observing the defect. The other parts of the network are said to be \emph{irrelevant} for observing the defect and can be removed. For example, for a defect caused by the algorithm trying to insert an XOR gate into an AIG, which is interpreted as inserting an AND gate instead, a core in the testcase may be a subnetwork computing the XOR function. 
Due to the localized-computation design style of modern scalable logic synthesis algorithms, the cores are usually small-sized.
We say that a core is \emph{minimal} if, for any node $n$, removing $n$ results in never observing the defect again no matter how the fanins and fanouts of $n$ are re-connected. A minimal core in a failure-inducing testcase may or may not be unique. 
The goal of testcase minimization is to find a minimal core in a given failure-inducing testcase.

We develop a testcase minimization tool for logic networks similar to delta debugging but without adopting binary search. Given a network and an AUT with verification (i.e. an oracle), our testcase minimizer iteratively tries to reduce the network and tests if the defect is still observed. Only the reduction operations that preserve observing the defect are kept; otherwise, the operation is undone. Different reduction operations are tried in six stages with increasing (finer) granularity as follows:

\begin{enumerate}[(a)]
    \item \emph{Remove PI}: Substitute a PI $n$ with constant zero, thus simplifying its TFO by constant propagation. In AIGs, some nodes in the TFO of $n$ that are connected to $n$ without complement are removed, and so are their MFFCs.
    \item \emph{Remove PO}: Substitute a PO $n$ with constant zero, thus removing its MFFC.
    \item \emph{Substitute gate}: Substitute a gate $n$ with constant zero, thus removing its MFFC and simplifying its TFO by constant propagation (as in (a)).
    \item \emph{Simplify TFO}: Assign fanins of a gate $n$ as new POs, and then substitute $n$ with constant zero. This operation is less aggressive than the previous one because only the TFO of $n$ is simplified and its MFFC is kept.
    \item \emph{Remove MFFC}: Substitute a gate $n$ with a new PI. This operation does not cause constant propagation in its TFO and only removes the MFFC of $n$.
    \item \emph{Remove gate}: Assign fanins of a gate $n$ as new POs, and then substitute $n$ with a new PI or with one of its fanins. Only $n$ is removed.
\end{enumerate}

\Cref{fig:stages} illustrates the effects of an operation in each of the reduction stages. Regions filled in blue are removed after the operation, and regions marked in yellow are simplified by constant propagation after the operation. Wires and PIs or POs drawn in green are added after the operation.

The relative granularity of stages \emph{remove PI} and \emph{remove PO} depends on the shape of the network. For networks with smaller TFO of PIs and less logic sharing in the TFI of POs, \emph{remove PO} reduces the network faster; for networks with smaller MFFC of POs and more reconvergences near the PIs, \emph{remove PI} reduces the network faster. Thus, the first stage to apply is heuristically decided by whether the network has more PIs than POs (\emph{remove PO} is applied first) or more POs than PIs (\emph{remove PI} is applied first).

In each stage, the minimizer backs-up the current network, randomly samples a PO or a gate as $n$ and performs the corresponding reduction operation. If the defect is not observed anymore after reduction, the back-up is restored. This procedure is repeated until all POs or all gates have been sampled, or until a pre-defined number of operations have been tried.

The resulting network after minimization cannot be reduced anymore because the last stage tries every possibility to remove one gate. Thus, by definition, the minimized testcase is guaranteed to be a minimal core. However, minimal cores are not necessarily unique, so it is possible that a different order of reduction operations (e.g. by using a different random seed) results in a smaller minimal core.

The minimized testcases are, in the most cases, highly destructed and cannot be recognized or reverse-engineered anymore. Therefore, the testcase minimizer does not only facilitate the debugging process, but also the communication between developers when commercially-sensitive benchmarks are involved. 

\subsection{Usage Example}\label{subsec:example}
The testing and debugging toolkit described in this section is implemented in \textit{mockturtle}\footnote{Available: \url{https://github.com/lsils/mockturtle}} as part of the EPFL open-source logic synthesis libraries~\cite{Soeken18}. The toolkit supports testing and debugging any application that takes a logic network, written in AIGER (for AIGs) or Verilog (for other network types supported in \emph{mockturtle}, such as XAGs and MIGs) formats, as input.

Figure~\ref{fig:ex} shows an example workflow of our toolkit. In this example, the toolkit is used to fuzz test an algorithm implemented in \emph{mockturtle} (marked in green), and then, if a defect is observed, minimizes the generated failure-inducing testcase (marked in red). This can be done similarly for other C++-based tools that include \emph{mockturtle} as a library.

\begin{figure}[ht]
\centering
\begin{lstlisting}[escapechar = $]
#include <mockturtle/mockturtle.hpp>
using namespace mockturtle;

int main()
{$\tikz[remember picture] \node [] (e){};$ 
  auto opt = [](aig_network aig) -> bool {
    aig_network const aig_copy = aig.clone();
    aig_resubstitution(aig);
    aig_network const miter = *miter(aig_copy, aig);
    return *equivalence_checking(miter);
  };
$\tikz[remember picture] \node [] (a){};$ 
  fuzz_tester_params fuzz_ps; 
  fuzz_ps.file_format = fuzz_tester_params::aiger;
  fuzz_ps.filename = "fuzz.aig";
  fuzz_ps.timeout = 20; // 20 minutes
  auto gen = random_aig_generator();
  network_fuzz_tester<aig_network, decltype(gen)> 
    fuzzer(gen, fuzz_ps);
  bool has_bug = fuzzer.run(opt);
$\tikz[remember picture] \node [] (b){};$ 
  if (!has_bug) return 0;
$\tikz[remember picture] \node [] (c){};$ 
  testcase_minimizer_params min_ps;
  min_ps.file_format = testcase_minimizer_params::aiger;
  min_ps.init_case = "fuzz";
  min_ps.minimized_case = "fuzz_min";
  testcase_minimizer<aig_network> minimizer(min_ps);
  minimizer.run(opt);
$\tikz[remember picture] \node [] (d){};$ 
  aig_network aig;
  lorina::read_aiger("fuzz_min.aig", aiger_reader(aig));
  write_dot(aig, "fuzz_min.dot");
  std::system("dot -Tpng -O fuzz_min.dot");

  return 0;
}
\end{lstlisting}

\begin{tikzpicture}[remember picture, overlay]
\draw[green!50!black,thick] ($(a.east)+(0,-0.1)$) rectangle ($(b)+(8.3,0)$);
\draw[red,thick] (c.east) rectangle ($(d)+(8.3,0)$);
\draw[blue,thick] ($(e.east)+(-0.15,-0.1)$) rectangle ($(a)+(8.3,0.1)$);

\draw[green!50!black,thick,fill=white,rounded corners] ($(a)+(6.35,0.05)$) rectangle ($(a)+(8.15,-0.3)$) node [midway, green!50!black] {Fuzzer};

\draw[red,thick,fill=white,rounded corners] ($(c)+(6.35,0.2)$) rectangle ($(c)+(8.15,-0.2)$) node [midway, red] {Minimizer};

\draw[blue,thick,fill=white,rounded corners] ($(e)+(6.2,0.1)$) rectangle ($(e)+(8.0,-0.3)$) node [midway, blue] {Oracle};

\end{tikzpicture} 
\vspace{-2em}
\caption{Example code to use the proposed toolkit to generate, minimize, and visualize a failure-inducing testcase.}\label{fig:ex}
\end{figure}

Our toolkit is also applicable for testing and debugging external tools. In this case, the lambda function in lines $6$ to $11$ in \Cref{fig:ex} shall be replaced by one resembles the code in \Cref{fig:ex-abc}.

\begin{figure}[ht]
\centering
\input{example_code2}
\vspace{-1em}
\caption{Example code to use the toolkit for testing and debugging an external tool, ABC.}\label{fig:ex-abc}
\vspace{-1em}
\end{figure}

Similar to \texttt{aigfuzz} and \texttt{aigdd}, calling the oracle as a system command requires switching the program control through the command shell and interfacing the testcases by reading and writing files. With the possibility of a tight integration as in \Cref{fig:ex}, these interfacing overheads can be eliminated and, empirically, making the automated testing and debugging workflow about $10\times$ faster.

\section{Case Study}\label{sec:case}
\begin{figure*}
\centering
\begin{subfigure}[b]{0.99\textwidth}
    \centering
    \includegraphics[width=\textwidth]{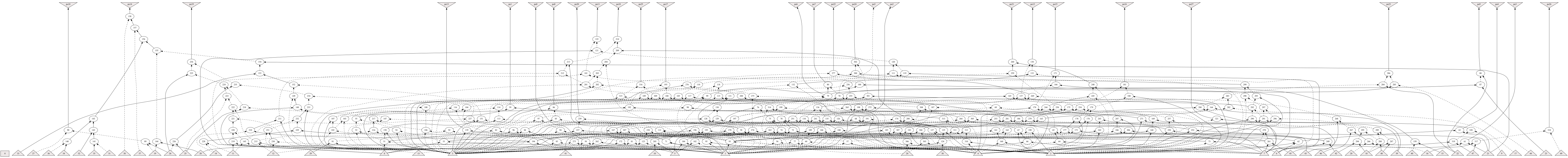}
    \caption{Before reduction, the original testcase is too big for human eyes to understand.}\label{fig:original}
\end{subfigure}
\hfill
\begin{subfigure}[t]{0.45\textwidth}
    \centering
    \includegraphics[width=\textwidth]{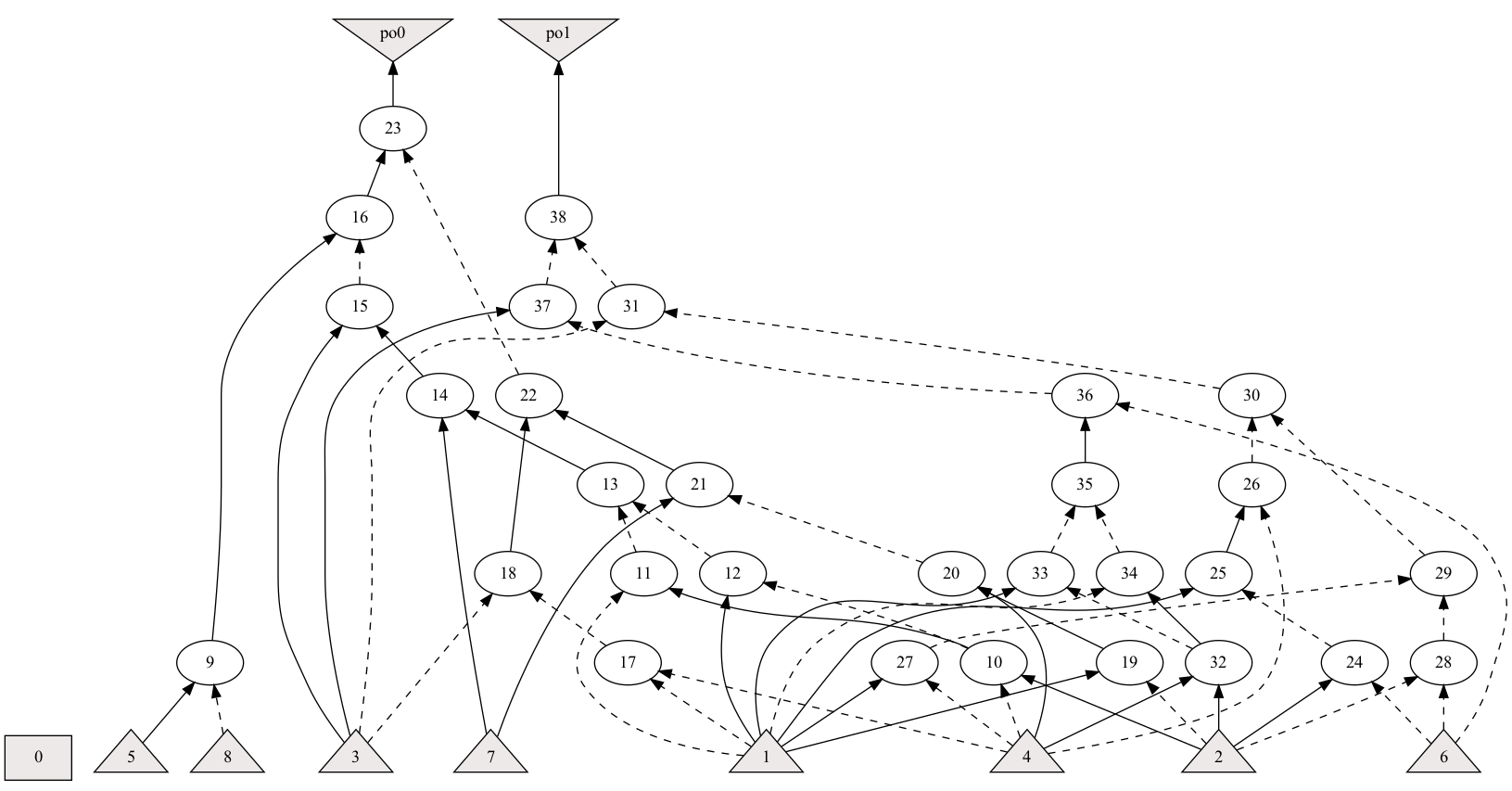}
    \caption{\textit{Remove PO} only.}\label{subfig:1}
\end{subfigure}
\hfill
\begin{subfigure}[t]{0.26\textwidth}
    \centering
    \includegraphics[width=\textwidth]{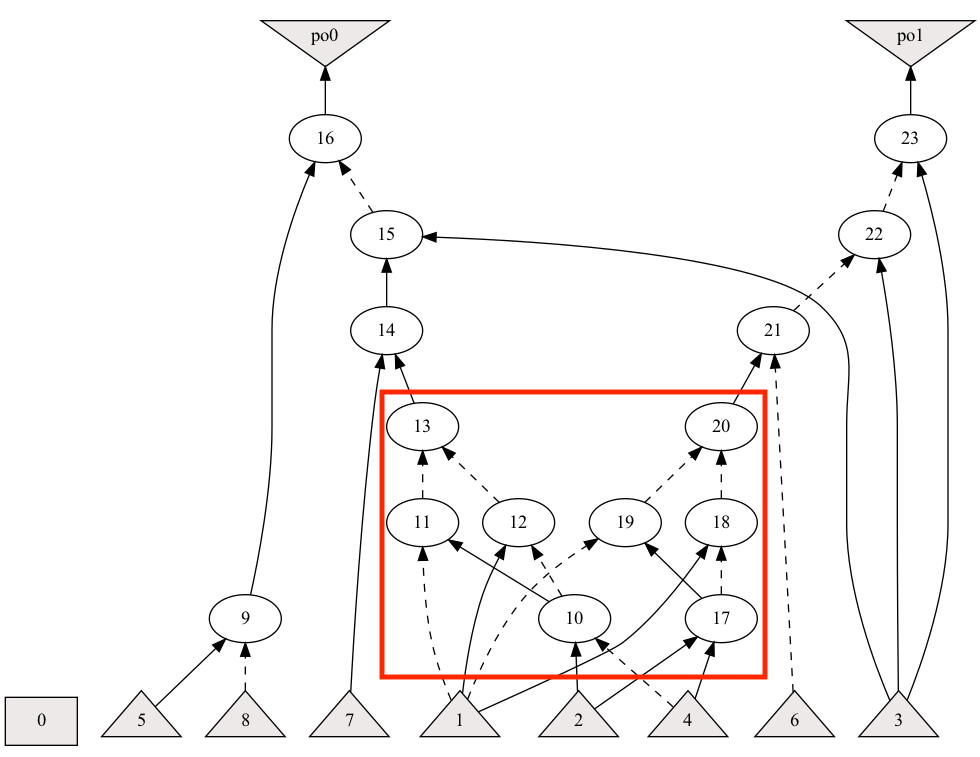}
    \caption{\textit{Remove PO} and \textit{substitute gate}.}\label{subfig:2}
\end{subfigure}
\hfill
\begin{subfigure}[t]{0.2\textwidth}
    \centering
    \includegraphics[width=\textwidth]{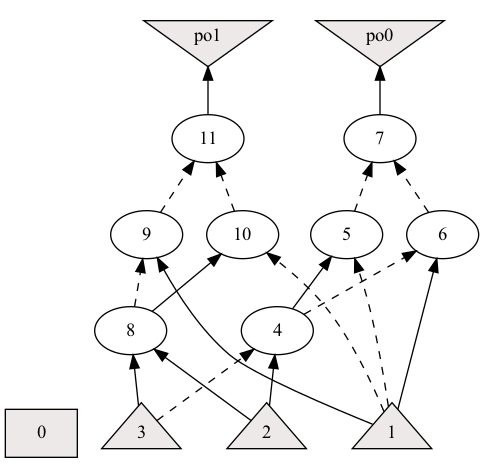}
    \caption{The minimum failure-inducing testcase.}\label{fig:minimum}
\end{subfigure}
\hfill
\vspace{-1em}
\caption{The failure-inducing testcase for an algorithm implemented in \emph{mockturtle} and intermediate results of minimizing it.}
\label{fig:case}
\end{figure*}

As a case study, we apply the toolkit on a known defect in a variation of \emph{cut rewriting}, which uses a compatibility graph to identify compatible substitution candidates~\cite{RienerHMMS19}, implemented in \emph{mockturtle}.\footnote{The function \texttt{cut\_rewriting\_with\_compatibility\_graph} can be found in \texttt{algorithms/cut\_rewriting.hpp}.} The defect can be observed by having a cyclic network after applying the algorithm. 
The failure-inducing core of this defect is shown in \Cref{fig:minimum}. The cyclic result is caused by the algorithm observing $n_7 \oplus n_2$ as a substitution for $n_{11}$ and $n_{11} \oplus n_2$ as a substitution for $n_7$, and trying to apply the two substitutions at the same time. To identify that the two substitution candidates are in conflict, the algorithm should check, for every pair $(A,B)$ of candidates, if the root of $A$ is contained in the cut of $B$ and the root of $B$ is contained in the cut of $A$. This would be a feasible fix for the defect, but would impact the efficiency of the algorithm. Another rewriting algorithm that does not use the compatibility graph but eagerly substitutes each candidate before searching for the next one is available in \emph{mockturtle}.\footnote{The function \texttt{cut\_rewriting} can be found in the same header file.} However, when not affected by the defect, the defective algorithm has on average better quality of result than eager rewriting. Also, the defect seems to be observed very rarely, as will be discussed in \Cref{subsec:case-fuzz}. As a compromise, both algorithms are kept in \emph{mockturtle}.

The first reported failure-inducing testcase for this defect is shown in \Cref{fig:original}. The original testcase was not minimized by the reporter and have $49$ PIs, $272$ AND gates, and $28$ POs. It took a human expert about $30$ minutes to manually reduce the testcase to \Cref{fig:minimum}, with $3$ PIs, $8$ gates and $2$ POs. Using the testcase minimizer, the original testcase is minimized to the same graph (subject to permutations of the two POs) within a second and using $94$ oracle calls. In \Cref{subsec:case-reduce}, we study the effectiveness and necessity of the reduction stages described in \Cref{subsec:minimization}.

\subsection{Capturing The Defect with Fuzz Testing}\label{subsec:case-fuzz}
\subsubsection{Using AIGs}
Knowing the existence of the defect, we investigate if our fuzz tester is capable of generating another failure-inducing testcase. However, even though the code line coverage has reached its maximum ($100\%$ excluding the lines disabled by the algorithm's options), the defect is not observed with more than a billion ($10^9$) regular (i.e., without leveraging knowledge of the known core) fuzz tests.
Even if we limit the sampling space to the $3$-input, $8$-gate, $2$-output topology as in \Cref{fig:minimum} and leaving only the connections to PIs and edge complementations as random choices, there are still $6^2 \times 3^4 \times 2^{16} = 191\,102\,976$ different possible networks, out of which only $3! \times 2^3 = 48$ networks (equivalent to \Cref{fig:minimum} subject to permutation and negation of PIs) are failure-inducing.

This case evidences that rare corner-case defects exist in logic synthesis applications, and the identification of them may only rely on real-world benchmarks. In these cases, the testcase minimization techniques are important to automatize the extraction of the failure-inducing core, which facilitates communication and debugging.

\subsubsection{Using XAGs}
We observe that the XOR functions in the core (nodes $9,10,11$ and nodes $5,6,7$ in \Cref{fig:minimum}) are necessary. Using any of the randomized methods described in \Cref{subsec:generation}, the possibility of generating an XOR function composed of three AIG nodes is low. However, it is much more likely to generate an XOR gate in an XAG. As the implementation is generic and works for both AIGs and XAGs, we can try to capture the defect using XAGs instead.
\Cref{tbl:fuzz-xag} shows that all the three methods successfully capture the defect within reasonable runtime.

\begin{table}[h]
\vspace{-0.3em}
\centering
\caption{Fuzzing the defective cut rewriting with XAGs.}
\label{tbl:fuzz-xag}
\vspace{-1em}
\begin{tabular}{lrr}
\toprule
Method & \#Tests & Time (s) \\
\midrule
Random & 8150 & 1.8 \\
Topology & 44498 & 6.6 \\
Composed & 77573 & 22.8 \\
\bottomrule
\end{tabular}
\vspace{-1em}
\end{table}
\vspace{-0.5em}

\subsection{Effects of The Reduction Stages}\label{subsec:case-reduce}
Given the initial failure-inducing testcase as in \Cref{fig:original}, using the default settings, our testcase minimizer produces the minimal failure-inducing testcase as in \Cref{fig:minimum}, which is a $97\%$ reduction rate in gate count. The minimality can be proved by trying to remove each gate and seeing that any possible resulting testcases are not failure-inducing. 

Figures \ref{subfig:1} and \ref{subfig:2} show the reduction results if only some reduction stages are applied. The first stage, \textit{remove PO} (\textit{remove PI} is skipped because there are more PIs than POs), provides already $89\%$ reduction of the testcase by removing large cones of irrelevant logic and quickly concentrates to the transitive fanin cone of two POs (\Cref{subfig:1}, $30$ gates). The next stage, \textit{substitute gate}, further reduces the size to $15$ gates (\Cref{subfig:2}), and the failure-inducing core is easily observable (marked with a red box). However, the other nodes on top of the core cannot be removed in this stage because substituting any of them with constant zero also removes part of the core. This can be accomplished by adding the stage \textit{simplify TFO}, resulting in \Cref{fig:minimum}. The two key operations are adding PO at nodes $13$ and $20$ and substituting nodes $14$ and $21$ with constant zero. It is also possible to reach the minimum by adding only the stage \textit{remove gate}, but it requires at least $6$ operations to remove nodes $14,15,16,21,22$ and $23$ one by one, showing that this stage operates in a finer granularity. It may seem that the stage \textit{remove MFFC} is not necessary. However, this is only because the failure-inducing core in this example does not have irrelevant transitive fanin gates (i.e., it is connected to PIs) in the original testcase. When this is not the case, the stages \textit{remove MFFC} and/or \textit{remove gate} are necessary to obtain the minimum.

\section{Experimental Results}\label{sec:exp}

\begin{table*}
\centering
\caption{Fuzz testing results.}
\label{tbl:fuzz}
\vspace{-1em}
\renewcommand{\arraystretch}{0.95}
\begin{tabular}{lrrrrrrrrrrrr}
\toprule

 & \multicolumn{3}{c}{\texttt{aigfuzz}} & \multicolumn{3}{c}{Random} & \multicolumn{3}{c}{Topology} & \multicolumn{3}{c}{Composed} \\ 
 & \multicolumn{3}{c}{\small\#Tests = $1000$} & \multicolumn{3}{c}{\small\#Tests = $1000$} & \multicolumn{3}{c}{\small\#Tests = $5000$} & \multicolumn{3}{c}{\small\#Tests = $5000$} \\ 
\cmidrule(lr){2-4}
\cmidrule(lr){5-7}
\cmidrule(lr){8-10}
\cmidrule(lr){11-13}

AUT & \#FITs & Size & Time & \#FITs & Size & Time & \#FITs & Size & Time & \#FITs & Size & Time \\
\midrule

\footnotesize{\texttt{mockturtle::aig\_resubstitution}} &
0 &     - & 10.3 &     0 &  - & 3.5 &    0 &      - & 0.02 &    1 & 14.0 & 0.03 \\
\footnotesize{\texttt{mockturtle::sim\_resubstitution}} & 
3 & 812.3 & 22.3 &     0 &  - & 4.2 &    6 & 5.0 & 0.06 &   93 & 21.7  & 0.11 \\

\hline
\footnotesize{\texttt{abc> bms\_start; if -u; strash}} &
952 & 2476.0 & 32.5 & 1000 & 950.0 & 14.6 &  1716 & 4.7 & 12.8 &  3749 & 20.3 & 13.0 \\
\footnotesize{\texttt{abc> \&if; \&mfs -dael; \&st}} &
956 & 2515.1 &  4.5 &  969 & 978.8 &  1.6 &     0 & - & 14.2 &  1047 & 23.9 & 12.3 \\
\footnotesize{\texttt{abc> \&mfsd; \&st}} &
473 & 3935.9 & 30.5 &  584 & 1078.4 & 14.1 &     0 & - & 14.1 &   120 & 24.2 & 14.1 \\
\footnotesize{\texttt{abc> \&mfsd -cd; \&st}} &
481 & 3959.7 &130.0 &   73 & 1266.4 & 47.8 &     0 & - & 14.2 &     1 & 20.0 & 14.4 \\
\footnotesize{\texttt{abc> if; mfse; strash}} & 
458 & 2763.3 & 14.7 &   93 & 940.3 & 13.6 &  1 & 5.0 & 16.0 &  1056 & 23.4 & 15.0 \\
\footnotesize{\texttt{abc> \&stochsyn resub}} & 
13 & 1164.8 & 14.6 &    0 &   - &  6.0 &     0 & - & 12.5 &    14 &  18.9 & 12.5 \\


\hline
\footnotesize{\texttt{lsoracle> aigscript}} &
 1 & 7364.0 & 38.1 &   0 &     - & 16.4 &     0 &     - & 32.8 &     1 & 14.0 & 32.8 \\
\footnotesize{\texttt{lsoracle> deep}} &
12 & 3227.3 & 41.6 &   0 &     - & 21.5 &     0 &     - & 33.3 &     6 & 23.0 & 32.8 \\
\footnotesize{\texttt{lsoracle> xmgscript}} &
 3 & 1056.0 & 21.6 &   2 & 350.0 & 10.5 &    38 & 4.9 & 11.5 &    99 &  20.1 & 12.0 \\
 
\midrule
Average & & 2941.6 & 32.8 &     & 995.5 & 14.0 &    & 4.7 & 14.7 &      & 21.5 & 14.5\\
\bottomrule
\end{tabular}
\end{table*}

\subsection{Fuzzing Open-Source Logic Synthesis Tools}\label{subsec:exp-fuzz}
To demonstrate the effectiveness of fuzzing and compare different testcase generation methods, we fuzz-tested the following open-source logic synthesis tools: \emph{mockturtle}\footnote{Available: \url{https://github.com/lsils/mockturtle}. Commit \texttt{cf4769f}.}~\cite{RienerTH+19}, \emph{ABC}\footnote{Available: \url{https://github.com/berkeley-abc/abc}. Commit \texttt{31519bd}.}~\cite{BraytonM10}, and \emph{LSOracle}\footnote{Available: \url{https://github.com/lnis-uofu/LSOracle}. Pull request \#81.}~\cite{NetoAT+19}. \Cref{tbl:fuzz} lists the commands or functions where defects have been observed. Fuzz testing campaigns were conducted on each AUT using \texttt{aigfuzz} and the three network generation methods described in \Cref{subsec:generation}. In each campaign, \texttt{aigfuzz} and the method Random ran $1000$ tests, whereas the methods Topology and Composed ran $5000$ tests.
In \Cref{tbl:fuzz}, column \#FITs lists the total number of failure-inducing testcases generated, column Size lists the average size (number of gates) of the failure-inducing testscases, and column Time lists the total runtime in minutes including the oracle calls. 

The Composed method captured defects in all of the listed AUTs. On average, Composed is about $2\times$ faster than \texttt{aigfuzz} and it tests on $5\times$ more networks. This is because the Composed testcases are, on average, $7\%$ in size comparing to those generated by \texttt{aigfuzz}. Also, notice that for AUTs in \emph{mockturtle}, the runtimes of our fuzzing methods are about $10\times$ faster than \texttt{aigfuzz} thanks to the tight integration.

\begin{table*}
\centering
\caption{Testcase minimization results.}
\label{tbl:minimize}
\vspace{-1em}
\renewcommand{\arraystretch}{0.95}
\begin{tabular}{lrrrrrrr}
\toprule

 & & \multicolumn{3}{c}{\texttt{aigdd}} & \multicolumn{3}{c}{Ours} \\
\cmidrule(lr){3-5}
\cmidrule(lr){6-8}

AUT & Original size & Size & \#Calls & Time & Size & \#Calls & Time \\
\midrule

\footnotesize{\texttt{mockturtle::cut\_rewriting\_with\_compatibility\_graph}} & 272 & 8 & 210 & 32.5 & 8 & 96 & 0.1 \\
\footnotesize{\texttt{mockturtle::sim\_resubstitution}} & 615 & 7 & 735 & 11.6 & 8 & 351 & 2.5 \\
\hline
\footnotesize{\texttt{abc> \&mfsd -cd; \&st}} & 1050 & 31 & 1198 & 150.0 & 20 & 857 & 120.5 \\
\footnotesize{\texttt{abc> if; mfse; strash}} & 1850 & 6 & 834 & 61.2 & 5 & 333 & 30.3 \\
\footnotesize{\texttt{abc> \&stochsyn resub}} & 3228 & 10 & 1124 & 59.6 & 8 & 411 & 21.1 \\

\bottomrule
\end{tabular}
\end{table*}

\subsection{Testcase Minimization}\label{subsec:exp-min}
We compare our testcase minimizer to \texttt{aigdd} using the user-reported failure-inducing testcase in \Cref{sec:case} and four bigger testcases found by fuzz testing in \Cref{subsec:exp-fuzz}. In \Cref{tbl:minimize}, column Size lists the number of gates of the original and the minimized testcases, column \#Calls lists the number of oracle calls and column Time lists the total runtime in seconds. It can be observed that our minimizer reduces the testcases into minimal cores of roughly the same or smaller sizes comparing to \texttt{aigdd}, using on average $50\%$ oracle calls and $50\%$ runtime.

\section{Conclusion and Discussion}
In this paper, we survey automated testing and debugging techniques and provide an open-sourced toolkit specialized for gate-level logic synthesis applications. While random fuzz testing can already catch many higher-frequency defects, the topology-based fuzzing methods provide a more systematic approach to thoroughly test topology-related corner cases. After failure-inducing testcases are found, the testcase minimizer can be used to reduce their size efficiently to facilitate manual debugging (and also anonymizing sensitive testcases). Moreover, our testcase minimization technique guarantees to find a minimal core in the failure-inducing testcase, which often gives insights on the cause of the defect and may also be used to categorize testcases for the same AUT.
The case study shows that (1) some defects may be difficult to catch by fuzz testing, thus testcase minimization is important when we need to rely on real-world testcases; and (2) testing with more functionally-compact networks, such as XAGs, may help to detect some defects in generic logic synthesis algorithms.
In the remaining of this section, we discuss more potentials of the toolkit and directions for future research.

\subsection{Non-deterministic Defects}
Non-deterministic defects may be hard to debug because they cannot always be reproduced. Non-determinism may come from a random number generator without a fixed seed, a race condition in concurrent computation, or accessing to uninitialized or unintended (index-out-of-bounds) memory. If a non-deterministic defect is first observed with a large testcase, it may be difficult to minimize it while maintaining the defect being observed. In such cases, fuzz testing may help generating smaller testcases to observe the defect deterministically.

\subsection{Other Applications of The Toolkit}
In addition to testing and debugging, the proposed tools can also be used for finding examples with specific properties. For example, an open problem in logic synthesis is whether it is better to heavily optimize an AIG before transforming into MIG, or to perform optimization directly with an MIG. Our toolkit can be used to generate minimal examples where one optimization script obtains better results than the other, which might help researchers identify weaknesses in the algorithms.

\subsection{Future Works}
Our network fuzzer currently does not support generating $k$-LUT networks easily without specifying all possible LUT functions as different gate types. This can be mitigated by integrating a random truth table generator.

In addition to minimizing the failure-inducing input networks, when the defective AUT involves multiple independent algorithms (i.e., a \emph{script} with a sequence of \emph{commands}), it would also be helpful to minimize the script and remove irrelevant commands. This can be accomplished by automatic binary search, similar to delta debugging.

\begin{acks}
We thank Max Austin for providing the original testcase discussed in \Cref{sec:case}. 
\end{acks}

\bibliographystyle{ACM-Reference-Format}
\bibliography{iwls}

\end{document}